\documentclass{ifacconf}
\usepackage{graphicx}      
\usepackage{natbib}        

\usepackage[utf8]{inputenc}
\usepackage{xcolor}
\usepackage{lipsum}
\usepackage{amsmath}
\usepackage{enumerate}
\usepackage{amsfonts}
\usepackage{mathtools}
\usepackage{siunitx}
\usepackage{float}
\usepackage{chemist}

\usepackage{tikz}
\usepackage{pgfplots}
\usetikzlibrary{shapes}
\usetikzlibrary{shapes.geometric, arrows}
\tikzstyle{block}=[rectangle, draw, fill=white, minimum size=2em]
\tikzstyle{arrow} = [thick,->,>=stealth]
\usetikzlibrary{calc}
\usetikzlibrary{automata}
\pgfplotsset{compat=1.16}
\definecolor{RWTHblau100}{rgb}{0.000000 0.329412 0.623529}

\usepackage[ruled,vlined,algo2e]{algorithm2e}

\SetCommentSty{mycommfont}
\SetArgSty{textnormal}

\definecolor{RWTHblau100}{rgb}{0.000000 0.329412 0.623529}
\definecolor{RWTHblau75}{rgb}{0.250980 0.498039 0.717647}
\definecolor{RWTHblau50}{rgb}{0.556863 0.729412 0.898039}
\definecolor{RWTHblau25}{rgb}{0.780392 0.866667 0.949020}
\definecolor{RWTHblau10}{rgb}{0.909804 0.945098 0.980392}
\definecolor{RWTHschwarz100}{rgb}{0.000000 0.000000 0.000000}
\definecolor{RWTHschwarz75}{rgb}{0.392157 0.396078 0.403922}
\definecolor{RWTHschwarz50}{rgb}{0.611765 0.619608 0.623529}
\definecolor{RWTHschwarz25}{rgb}{0.811765 0.819608 0.823529}
\definecolor{RWTHschwarz10}{rgb}{0.925490 0.929412 0.929412}
\definecolor{RWTHmagenta100}{rgb}{0.890196 0.000000 0.400000}
\definecolor{RWTHmagenta75}{rgb}{0.913725 0.376471 0.533333}
\definecolor{RWTHmagenta50}{rgb}{0.945098 0.619608 0.694118}
\definecolor{RWTHmagenta25}{rgb}{0.976471 0.823529 0.854902}
\definecolor{RWTHmagenta10}{rgb}{0.992157 0.933333 0.941176}
\definecolor{RWTHyellow100}{rgb}{1.000000 0.929412 0.000000}
\definecolor{RWTHyellow75}{rgb}{1.000000 0.941176 0.333333}
\definecolor{RWTHyellow50}{rgb}{1.000000 0.960784 0.607843}
\definecolor{RWTHyellow25}{rgb}{1.000000 0.980392 0.819608}
\definecolor{RWTHyellow10}{rgb}{1.000000 0.992157 0.933333}
\definecolor{RWTHpetrol100}{rgb}{0.000000 0.380392 0.396078}
\definecolor{RWTHpetrol75}{rgb}{0.176471 0.498039 0.513725}
\definecolor{RWTHpetrol50}{rgb}{0.490196 0.643137 0.654902}
\definecolor{RWTHpetrol25}{rgb}{0.749020 0.815686 0.819608}
\definecolor{RWTHpetrol10}{rgb}{0.901961 0.925490 0.925490}
\definecolor{RWTHtuerkis100}{rgb}{0.000000 0.596078 0.631373}
\definecolor{RWTHtuerkis75}{rgb}{0.000000 0.694118 0.717647}
\definecolor{RWTHtuerkis50}{rgb}{0.537255 0.800000 0.811765}
\definecolor{RWTHtuerkis25}{rgb}{0.792157 0.905882 0.905882}
\definecolor{RWTHtuerkis10}{rgb}{0.921569 0.964706 0.964706}
\definecolor{RWTHgruen100}{rgb}{0.341176 0.670588 0.152941}
\definecolor{RWTHgruen75}{rgb}{0.552941 0.752941 0.376471}
\definecolor{RWTHgruen50}{rgb}{0.721569 0.839216 0.596078}
\definecolor{RWTHgruen25}{rgb}{0.866667 0.921569 0.807843}
\definecolor{RWTHgruen10}{rgb}{0.949020 0.968627 0.925490}
\definecolor{RWTHmaigruen100}{rgb}{0.741176 0.803922 0.000000}
\definecolor{RWTHmaigruen75}{rgb}{0.815686 0.850980 0.360784}
\definecolor{RWTHmaigruen50}{rgb}{0.878431 0.901961 0.603922}
\definecolor{RWTHmaigruen25}{rgb}{0.941176 0.952941 0.815686}
\definecolor{RWTHmaigruen10}{rgb}{0.976471 0.980392 0.929412}
\definecolor{RWTHorange100}{rgb}{0.964706 0.658824 0.000000}
\definecolor{RWTHorange75}{rgb}{0.980392 0.745098 0.313725}
\definecolor{RWTHorange50}{rgb}{0.992157 0.831373 0.560784}
\definecolor{RWTHorange25}{rgb}{0.996078 0.917647 0.788235}
\definecolor{RWTHorange10}{rgb}{1.000000 0.968627 0.917647}
\definecolor{RWTHrot100}{rgb}{0.800000 0.027451 0.117647}
\definecolor{RWTHrot75}{rgb}{0.847059 0.360784 0.254902}
\definecolor{RWTHrot50}{rgb}{0.901961 0.588235 0.474510}
\definecolor{RWTHrot25}{rgb}{0.952941 0.803922 0.733333}
\definecolor{RWTHrot10}{rgb}{0.980392 0.921569 0.890196}
\definecolor{RWTHbordeaux100}{rgb}{0.631373 0.062745 0.207843}
\definecolor{RWTHbordeaux75}{rgb}{0.713725 0.321569 0.337255}
\definecolor{RWTHbordeaux50}{rgb}{0.803922 0.545098 0.529412}
\definecolor{RWTHbordeaux25}{rgb}{0.898039 0.772549 0.752941}
\definecolor{RWTHbordeaux10}{rgb}{0.960784 0.909804 0.898039}
\definecolor{RWTHviolett100}{rgb}{0.380392 0.129412 0.345098}
\definecolor{RWTHviolett75}{rgb}{0.513725 0.305882 0.458824}
\definecolor{RWTHviolett50}{rgb}{0.658824 0.521569 0.619608}
\definecolor{RWTHviolett25}{rgb}{0.823529 0.752941 0.803922}
\definecolor{RWTHviolett10}{rgb}{0.929412 0.898039 0.917647}
\definecolor{RWTHlila100}{rgb}{0.478431 0.435294 0.674510}
\definecolor{RWTHlila75}{rgb}{0.607843 0.568627 0.756863}
\definecolor{RWTHlila50}{rgb}{0.737255 0.709804 0.843137}
\definecolor{RWTHlila25}{rgb}{0.870588 0.854902 0.921569}
\definecolor{RWTHlila10}{rgb}{0.949020 0.941176 0.968627}
\definecolor{rwthblue}{rgb}{0,0.4,0.8}    
\definecolor{hellgrau}{rgb}{0.9,0.9,0.9}     

\newcolumntype{P}[1]{>{\centering\arraybackslash}p{#1}}
\newcommand\Tstrut{\rule{0pt}{2.6ex}}         
\newcommand\Bstrut{\rule[-0.9ex]{0pt}{0pt}}   

\begin{document}




\begin{frontmatter}
\title{Complementarity-constrained predictive control for efficient gas-balanced \\hybrid power systems\thanksref{footnoteinfo}} 
%

%
\thanks[footnoteinfo]{The Research Council of Norway funds this research through PETROSENTER LowEmission (project code 296207).}

\author[NTNU]{Kiet Tuan Hoang} 
\author[SINTEF]{Brage Rugstad Knudsen} 
\author[NTNU]{Lars Struen Imsland}

\address[NTNU]{Department of Engineering Cybernetics, Norwegian University of Science and Technology, Trondheim, Norway. E-mail:   kiet.t.hoang@ntnu.no, lars.imsland@ntnu.no}
\address[SINTEF]{SINTEF Energy Research AS, Trondheim, Norway. E-mail:   
brage.knudsen@sintef.no}
\begin{abstract}
Controlling gas turbines (GTs) efficiently is vital as GTs are used to balance power in onshore/offshore hybrid power systems with variable renewable energy and energy storage. However, predictive control of GTs is non-trivial when formulated as a dynamic optimisation problem due to the semi-continuous operating regions of GTs, which must be included to ensure complete combustion and high fuel efficiency. This paper studies two approaches for handling the semi-continuous operating regions of GTs in hybrid power systems through predictive control, dynamic optimisation, and complementarity constraints. The proposed solutions are qualitatively investigated and compared with baseline controllers in a case study involving GTs, offshore wind, and batteries. While one of the baseline controllers considers fuel efficiency, it employs a continuous formulation, which results in lower efficiency than the two proposed approaches as it does not account for the semi-continuous operating regions of each GT. 
\end{abstract}

\end{frontmatter}

\maketitle

\section{Introduction}

\begin{figure}
    \centering
    {\includegraphics[width=0.95\columnwidth, trim = 0.4cm 4.9cm 9.cm 3.4cm, clip]{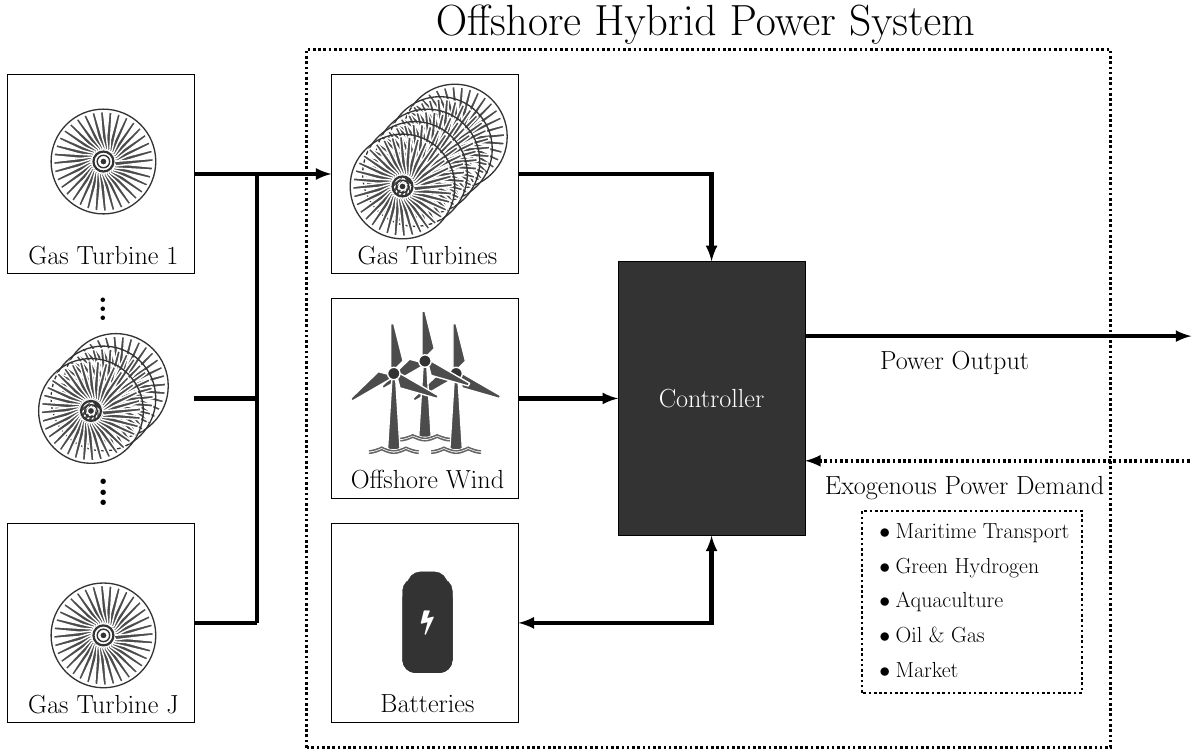}}
    \caption{An illustration of offshore hybrid power systems.}
    \label{fig:1-offshorehybridpowersystemSchematic}
\end{figure}

A large share of Norway's greenhouse gas (GHG) emissions stem from using offshore gas turbines (GTs) as the primary power source \citep{NPD:2019-ResourceReport}. In generating electricity, fossil fuel is combusted, thus releasing GHG emissions. To reduce the emissions, alternative power sources such as offshore wind with batteries can be included in the grid to form offshore hybrid power systems (OHPS) to increase the penetration of renewable energy, see Fig \ref{fig:1-offshorehybridpowersystemSchematic}. Such a modification to the offshore energy infrastructure can be achieved without modifying the operation of the GTs, as the reduction in emissions comes from replacing part of the GT power through renewable generation \citep{Hoang:2022-NMPC-OHPS}. Besides increasing the penetration of renewable energy, operating GTs more efficiently with predictive control reduces emissions \citep{Hoang:2023-M-ENMPC-OHPS}. Such approaches, though, are limited as GTs should generally be operated at all but below minimum load or during start-up to ensure complete combustion and high fuel efficiency \citep{GE:2001-GT-emissions-and-control}. This results in semi-continuous formulations which can be handled via mixed-integer variables \citep{7112194, JIANG2021106460} to ensure complete combustion while enabling the GTs to turn themselves off when not needed. However, this also results in nonlinear mixed-integer formulations, which may be intractable if the prediction horizon for which the nonlinear dynamic optimisation problems aims to solve is too large \citep{Koppe:2012-Complexit-of-Nonlinear-Mixed-integer}. For OHPS, these prediction horizons may be in hours or even days due to the use of weather forecasts for increased performance \citep{Hoang:2023-M-ENMPC-OHPS}. 

This paper proposes to use complementarity constraints to approximate the discrete formulation with continuous formulations. The idea is that given some complementarity constraints, a continuous variable can be formulated to exhibit the same discrete behaviour as a discrete variable, which can be solved with standard nonlinear program solvers \citep{Baumrucker:2008-MPEC-Problem-formulations,Biegler:2010-NLP-book}. Based on complementarity constraints, this paper proposes two different economic nonlinear model predictive control (ENMPC) formulations to account for the semi-continuous operating region of GTs to operate the GTs efficiently in an OHPS case study with variable wind and batteries.

The paper starts with section \ref{sec:2-system} to present the OHPS with a focus on GTs. Section \ref{sec:3-method} presents the ENMPC formulations without and with complementarity constraints. The methods are then validated in simulation with Section \ref{sec:4-results}.

\section{Offshore hybrid power systems}\label{sec:2-system}
This section aims to give an overview of the OHPS and focus on how GTs with varying maximum power outputs can be operated efficiently. 

\subsection{Modeling of Offshore Hybrid Power System}\label{subsec:2-system-summary}

This paper considers an isolated nonlinear OHPS consisting of a GT cluster with J different gas turbine generator (GTG) systems with varying maximum power output, a wind turbine generator (WTG) system based on a static power curve, and a battery (Bat) system with the following system dynamics
${f}: \mathbb{R}^{n_{x}}\!\times\mathbb{R}^{n_{u}}\!\times\mathbb{R}^{n_{p}}\!\rightarrow\mathbb{R}^{n_{x}}$ described by the system state ${x}(t)\!\in\mathbb{R}^{n_{x}}$, input ${u}(t)\!\in\mathbb{R}^{n_{u}}$, and parameter ${p}(t)\!\in\mathbb{R}^{n_{p}}$ \vspace*{-3 mm}

\begin{equation} 
    \begin{aligned}
    \dot{x} &= {f\left( x(t),u(t),p(t)\right)},
    \end{aligned}\label{eq:2-ohps-model-process} 
\end{equation}

where ${x}(t) =  \left[V^\text{gtg}_j(t) \ P^\text{gtg}_j(t) \ \hdots \ \text{SOC}^\text{bat}(t)\right]^\top $, ${u}(t) = \left[T^\text{gtg}_j(t)\ \hdots \ I^\text{bat}(t)\right]^\top$, and ${p}(t) = \left[\text{v}^\text{wind}(t) \ P^\text{wtg}(t)\right]^\top$ in which $V^\text{gtg}_{j}(t)$ [pu] and $P^\text{gtg}_{j}(t)$ [\si{\watt}] describe the GTG fuel flow and power output where ${j} \in [1,\text{J}]$, $\text{SOC}^\text{bat}(t)$ [\%] describes the battery state of charge, $T^\text{gtg}(t)$ [pu] describes the GTG throttle, $I^\text{bat}(t)$ [\si{\A}] describes the battery current, $\text{v}^\text{wind}(t)$ [\si{\metre\per\second}] describes the average WTG rotor wind speed, and $P^\text{wtg}(t)$ [\si{\watt}] describes the WTG power output (refer to \citet{Hoang:2022-NMPC-OHPS} and the references therein for further information on the modelling and assumptions).

\subsection{Control Objective}\label{subsec:2-control-obj}

Generally, the control objective of this OHPS is to, in decreasing order of importance \citep{Hoang:2023-M-ENMPC-OHPS} 
\begin{enumerate}
    \item \label{itm:first_goal} Satisfy the uncertain total power demand
    \item \label{itm:second_goal} Maximise WTG power to reduce GHG emissions
    \item \label{itm:third_goal} Optimal GTG operation to reduce emitted $\text{CO}_2$
    \item \label{itm:fourth_goal} Maximise the battery SOC for system flexibility 
    \item \label{itm:fifth_goal} Minimise actuator effort
\end{enumerate}
This paper will focus on control objective \ref{itm:third_goal} and highlight the impact of introducing semi-continuous optimisation variables in the controller while assuming that control objectives \ref{itm:first_goal}, \ref{itm:second_goal}, \ref{itm:fourth_goal}, and \ref{itm:fifth_goal} are satisfied sufficiently. 

\begin{figure}[t]
    \centering
    {\includegraphics[width=0.95\columnwidth, trim = 1.1cm 0cm 2cm 2.5cm, clip]{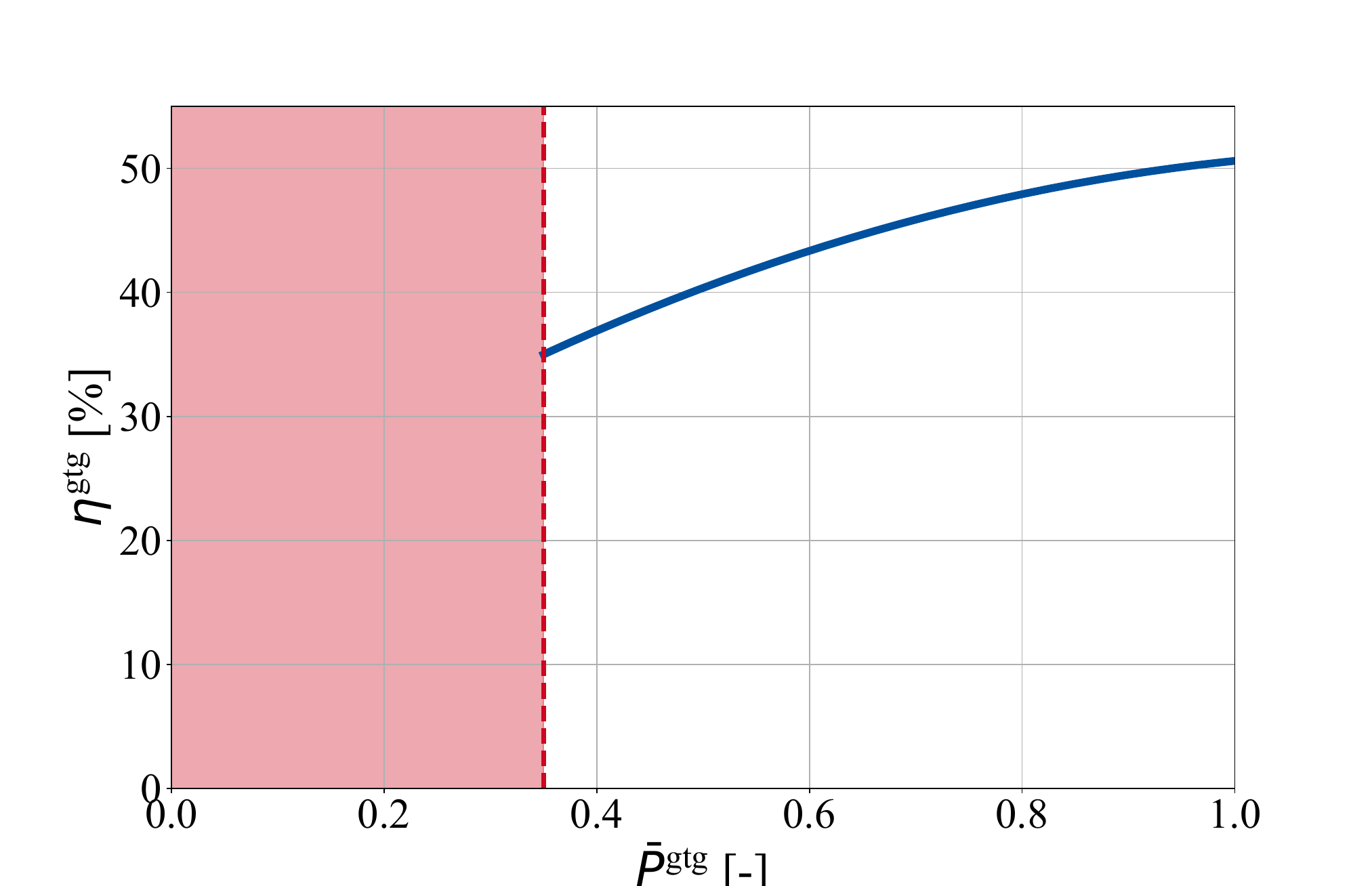}}
    \caption{The relationship between GTG efficiency $\eta^\text{gtg}$ and partial load as defined by \eqref{eq:2-gtg-efficiency} and \eqref{eq:2-gtg-partial}. A cut-off at 35\% partial load based on \cite{GE:2022-factsheet} in red illustrates the semi-continuous operating region of GTGs.}
    \label{fig:2-GTG-efficiency}
\end{figure}

Optimal GTG operation can mean many things. This paper focuses on high fuel efficiency, as the fuel efficiency $\eta^\text{gtg}(t)$ of each GTG directly affects the emitted GHG emissions (measured in CO$_2$) \vspace*{-3 mm} 

\begin{align}
        \dot{m}^{\text{CO}_2}(t) = \frac{\text{M}^\text{CO$_{2}$}}{\text{M}^\text{CH$_{4}$}}\dot{m}^{\text{CH}_4}(t) = \frac{\text{M}^\text{CO$_{2}$}}{\text{M}^\text{CH$_{4}$}} \frac{P^\text{gtg}(t)}{\eta^\text{gtg}(t)\text{LHV}^{\text{CH}_{{4}}}}.\label{eq:2-CO2-GHG}
\end{align}

Equation \ref{eq:2-CO2-GHG} is derived by assuming ideal stoichiometric combustion of methane with air\vspace*{-3 mm} 

\begin{chemmath}
  CH_{4}+2(O_{2}+3.76N_{2})
  \reactrarrow{0pt}{0.5cm}{}{}
  CO_{2} + 2H_{2}O + 7.52N_{2},
\end{chemmath}

where M is the molecular weight [g/mole], $\dot{m}(t)$ [kg/s] is the mass flow of reactant/product, LHV is the lower heating value, and where the fuel efficiency $\eta^\text{gtg}(t)$ can be approximated and computed from a semi-continous second order efficiency curve based on a normalised minimum load requirement $\bar{P}^\text{gtg,min}$ \citep{Nirbito:2020-GTG-performance-Boat} \vspace*{-3 mm} 

\begin{equation}
  \eta^\text{gtg}(t)=\begin{cases}
     \alpha_1 {\bar{P}^\text{gtg}(t)}^2 + \alpha_2 \bar{P}^\text{gtg}(t), & \text{if $\bar{P}^\text{gtg}(t)\geq\bar{P}^\text{gtg,min}$} .\\
     0, & \text{otherwise}\label{eq:2-gtg-efficiency},
  \end{cases}
\end{equation}

where $\alpha_i$ are fitted constants and $\bar{P}^\text{gtg}(t)$ is the normalised power based on the maximum output ${P}^\text{gtg,max}$\vspace*{-3 mm} 

\begin{equation}
    \bar{P}^\text{gtg}(t) = \frac{P^\text{gtg}(t)}{P^\text{gtg,max}}.\label{eq:2-gtg-partial}
\end{equation}

As can be derived from \eqref{eq:2-CO2-GHG}, \eqref{eq:2-gtg-efficiency}, \eqref{eq:2-gtg-partial}, and Fig. \ref{fig:2-GTG-efficiency}, optimal operation of GTGs despite the semi-continuous formulation for $\eta^\text{gtg}(t)$ can be achieved by continuously maximising $\eta^\text{gtg}(t)$ for each of the GTGs $\text{if $\bar{P}^\text{gtg}(t)\geq\bar{P}^\text{gtg,min}$}$ and not include the possibility of shutting down each GTG in the dynamic optimisation problem. A disadvantage of such an approach is that it limits how multiple GTGs can be operated, as there are scenarios where shutting down one or more GTGs can improve $\eta^\text{gtg}(t)$ as GTGs are inefficient at low partial load. Based on this observation, this study aims to show how this limitation can be handled by enabling the semi-continuous operating regions of each GTG in the dynamic optimisation problem with the introduction of complementarity constraints. Two approaches are subsequently shown which optimise $\eta^\text{gtg}(t)$

\begin{enumerate}
    \item \textbf{Directly} by including $\eta^\text{gtg}(t)$ in the cost function. The resulting controller can thus focus on maximising $\eta^\text{gtg}(t)$ by operating each GTG at a high load and possibly completely shutting down each GTG. 
    \item \textbf{Indirectly} by only allowing each GTG to be at maximum load or completely shut down. The resulting controller prevents each GTG from operating at partial load, which is characterised by lower efficiency.
\end{enumerate}

\section{Optimal control of gas turbines}\label{sec:3-method}

This section describes the two proposed approaches for maximising fuel efficiency for a cluster of GTGs in an OHPS with complementarity-constrained ENMPC. First, a baseline ENMPC approach which does not consider the semi-continuous operating region of the GTGs is presented in Section~\ref{subsec:3-1-ENMPC} before the two approaches with complementarity constraints are given in Sections~\ref{subsec:3-2-Mixed-ENMPC1} (direct approach) and ~\ref{subsec:3-3-Mixed-ENMPC2}  (indirect approach).

\subsection{Control with economic cost}\label{subsec:3-1-ENMPC}

One way of optimal operation of OHPSs without considering the semi-continous operating region of GTGs is to utilise an ENMPC, given some constraints and an economic cost $\text{V}\left(x(t),u(t)\right)$ \citep{Rawlings:2017-MPCBook}\vspace*{-3 mm} 

\begin{equation}
\begin{alignedat}{2}
&\! \min_{{x(t),u(t)}}        &\quad& \int_{t_0}^{t_0+N_P}\text{V}\left(x(t),u(t)\right)\mathrm{d}t, \\
&\text{s.t} &      & {{\dot{x}} =  f\left(x(t),u(t),p(t)\right)}, \ t \in [t_0 , N_P]\\
&                  &      & 0 = g_P\left(x(t),u(t),p(t)\right), \ t \in [t_0 , N_P]\\
&                  &      & 
h\left(x(t),u(t),p(t)\right), \ t \in [t_0 , N_P]\\
&                  &      & {x}(t) \in \mathbb{X}\subseteq \mathbb{R}^{n_{x}}, \ t \in [t_0 , N_P]\\
&                  &      & {u}(t) \in \mathbb{U} \subseteq \mathbb{R}^{n_{u}}, \ t \in [t_0 , N_P]\\
&                  &      & {x}(t_0)={x}_{t_0},
\end{alignedat}\label{eq:3-NLP-Formulation-General}
\end{equation}

to maximise the efficiency and minimise actuator effort\vspace*{-3 mm} 

\begin{align}
    \text{V}(x(t),u(t)) =-\sum_{{j}=1}^\text{J}K_j^\text{gtg} \eta_j^\text{gtg}(t) + u(t)^\top K^u_1 u(t),\label{eq:3-cost-function-ENMPC}
\end{align}

with $K$ being positive matrices/constants to be tuned and \vspace*{-6 mm} 

\begin{align}
    g_P(x(t),u(t),p(t)) = &\sum_{{j}=1}^\text{J}  P_{j}^\text{gtg}(t) + P^\text{wtg}(t)\label{eq:3-ENMPC-power-demand}\\&+P^\text{bat}(t) -   P^\text{demand}(t),\nonumber
\end{align}

which boils down to solving a finite-horizon nonlinear program (NLP) with standard techniques where $N_P$ is the prediction horizon, $\dot{x} = {f(\cdots)}$ are dynamic model constraints, ${g}_P(\cdots)$ is a constraint to ensure that the power demand $P^\text{demand}(t)$ is always satisfied, and ${x}(t)$ and ${u}(t)$ are the decision variables to be optimised for the prediction horizon inside of some set $\mathbb{X}$ and $\mathbb{U}$ where ${u}(t_0 + \mathrm{d}t)$ is the optimal control policy when solving \eqref{eq:3-NLP-Formulation-General}. The control law is computed iteratively in a receding horizon manner at every time step $\mathrm{d}t$ based on the current initial system state $x_{t_0}$ while adhering to the operating region of each GTG specified by the minimum and maximum load requirement \vspace*{-3 mm} 

\begin{equation}
    h(x(t),u(t),p(t)) = P^\text{gtg,min} \leq P^\text{gtg}(t) \leq P^\text{gtg,max}.\label{eq:3-hlc-inequality_const}    
\end{equation}

Since the GTG operates within these requirements, $\eta^\text{gtg}(t)$ can be optimised without introducing discrete variables. However, this also means that the GTGs cannot be turned off as the ENMPC does not consider the whole semi-continuous operating region of GTGs. It is essential to note that generally, having the GT with the largest output at high efficiency is more critical, as that GT burns more fuel than the smaller GTs, thus releasing more GHG.  This can be achieved by letting $K_1^\text{gtg}(t)>K_2^\text{gtg}(t)......>>K_\text{J}^\text{gtg}(t)$ when tuning \eqref{eq:3-cost-function-ENMPC}.

\subsection{Direct control with complementarity constraints}\label{subsec:3-2-Mixed-ENMPC1}

A mixed-integer formulation can be used to account for the semi-continuous operation region of each GTG. To include shut-down flexibility in the optimisation problem, the following inequality constraint can be used with a binary variable $z_j(t)$ for each of the GTGs \vspace*{-3 mm} 

\begin{align}
   z_j(t)P_j^\text{gtg,min}\leq{P_{j}^\text{gtg}}(t)\leq z_j(t)P_j^\text{gtg,max} ,\ j\in [1, \text{J}],\label{eq:3-first-constraint-complementarity-constrained-approach}
\end{align}

to impose the following logic \vspace*{-3 mm} 

\begin{equation}
  {P}_j^\text{gtg}(t)\in\begin{cases}
     \left[{P}_j^\text{gtg,min},{P}_j^\text{gtg,max}\right], & \text{if} \ z_j(t) = 1 .\\
    0, &  \text{if} \ z_j(t) = 0.\label{eq:3-Decision_logic_first}
  \end{cases}
\end{equation}

To approximate $z_j(t)$, two continous variables $y_j(t)\in[0,1]$ and $q_j(t)\in[0,1]$ are introduced and approximated to be binary with the following constraint \citep{Baumrucker:2008-MPEC-Problem-formulations}\vspace*{-3 mm} 

\begin{equation}
    0\leq {y_j(t)}{q_j(t)}\leq 0\label{eq:3-complementarity-for-gtg},
\end{equation}

which are made complementary with \vspace*{-3 mm} 

\begin{equation}
    q_j(t) = 1-y_j(t).\label{eq:3-aux-variable-constraint}
\end{equation}

Equations \eqref{eq:3-complementarity-for-gtg} and \eqref{eq:3-aux-variable-constraint} forces $y_j(t)$ and $q_j(t)$ to be either 0 and 1. A new ENMPC formulation can then be defined with complementarity constraints to directly optimise $\eta^\text{gtg}_j(t)$ with the same structure as the ENMPC from \eqref{eq:3-NLP-Formulation-General} where the complementarity constraints are enforced by replacing \eqref{eq:3-hlc-inequality_const} with \eqref{eq:3-first-constraint-complementarity-constrained-approach}, \eqref{eq:3-complementarity-for-gtg}, and \eqref{eq:3-aux-variable-constraint} for all J GTGs. To avoid infeasibility and numerical stability, the complementarity constraints have been implemented as a penalty term with a regularisation term $\mathrm{d}y_j(t) = y_j(t) - y_j(t+\mathrm{d}t)$ in the cost instead. By penalising $\mathrm{d}y_j(t)$, the controller with complementarity constraints is regularised to minimise excessive switching between 0 and 1 for the new variables. The new cost function for the proposed complementarity-constrained ENMPC, which takes into account shut-down and start-up, can be defined according to \eqref{eq:3-cost-function-proposed-ENMPC-first} by modifying the previous ENMPC formulation \eqref{eq:3-NLP-Formulation-General} with a modified cost function which enables the GTGs to be optimised over the whole semi-continuous operating region. \vspace*{-3 mm}

\begin{align}
    \text{V}^\text{cc}_1\left(x(t),u(t),y(t),q(t)\right) =&\sum_{{j}=1}^\text{J}(\mathrm{d}y_j(t)^\top K_{j,1}^{\mathrm{d}y} \mathrm{d}y_j(t) + \label{eq:3-cost-function-proposed-ENMPC-first} \\ & \ K_{j,1}^{y}{y}_j(t){q}_j(t))  + \text{V}(x(t),u(t)).\nonumber
\end{align}

\subsection{Indirect control with complementarity constraints}\label{subsec:3-3-Mixed-ENMPC2}

Another way of using complementarity-constrained ENMPC to control GTGs is by preventing each GTG from ever being at partial load, i.e., always at either maximum partial load or turned off\vspace*{-3 mm} 

\begin{equation}
  {P}_j^\text{gtg}(t)=\begin{cases}
     {P}_j^\text{gtg,max}, & \text{if} \ y_j(t) = 1 .\\
    0, &  \text{if} \ y_j(t) = 0.\label{eq:3-Decision_logic_second}
  \end{cases}
\end{equation}

Such an approach is efficient, as each GTG is the most efficient at its maximum partial load with no emission released when turned off. Similar to \eqref{eq:3-first-constraint-complementarity-constrained-approach}, an equality constraint for each GTG can be formulated to ensure this bang-bang behaviour   \vspace*{-3 mm} 

\begin{align}
    g_2(x(t),y(t))=& {P_{j}^\text{gtg}}(t)-{y}_{j}(t){P}^\text{gtg,max}_{j} ,\ j\in [1, \text{J}].\label{eq:3-gas-turbine-power-linear-regression}
\end{align}

A new ENMPC formulation can then be defined with complementarity constraints again with the same structure as the ENMPC from \eqref{eq:3-NLP-Formulation-General} where the complementarity constraints for indirect optimal control are enforced by replacing \eqref{eq:3-first-constraint-complementarity-constrained-approach} with \eqref{eq:3-gas-turbine-power-linear-regression} with a new cost function which now omits $\eta^\text{gtg}(t)$ all together \vspace*{-3 mm} 

\begin{align}
    \text{V}^\text{cc}_2(x(t),u(t),y(t),q(t)) =&\sum_{{j}=1}^\text{J}(\mathrm{d}y_j(t)^\top K_{j,2}^{\mathrm{d}y} \mathrm{d}y_j(t) + \label{eq:3-cost-function-proposed-ENMPC-second} \\ & \ K_{j,2}^{y}{y}_j(t){q}_j(t))  + u(t)^\top K^u_2 u(t).\nonumber
\end{align}

It is important to notice from the last ENMPC formulation that $\eta^\text{gtg}(t)$ is not directly optimised in the dynamic optimisation problem but rather indirectly optimised. With the last formulation, prioritising the largest GTG for high fuel efficiency is not required, as one can assume that given fast enough switching, $\eta^\text{gtg}(t)$ will always be for all GTGs at its maximum fuel efficiency due to the equality constraint from \eqref{eq:3-gas-turbine-power-linear-regression} since $\bar{P}^\text{gtg}(t) \notin \left(0,\bar{P}^\text{gtg,max}\right)$ where the efficiency is sub-optimal, see \eqref{eq:2-gtg-efficiency} and Fig. \ref{fig:2-GTG-efficiency}. This approach bypasses the minimum load requirement as $\bar{P}^\text{gtg}(t) \notin \left(0,\bar{P}^\text{gtg,min}\right)$.

\section{Simulation and Validation}\label{sec:4-results}
This section aims to show how the proposed complementarity-constrained ENMPCs can be used to control GTGs in an OHPS for reference tracking. For comparison reasons, the previous methods are compared. First, a baseline ENMPC which follows \eqref{eq:3-NLP-Formulation-General} is investigated where $\eta^\text{gtg}(t)$ is directly optimised, but where the semi-discrete behaviour of GTGs is not accounted for. Thus, the GTGs can only operate above the minimum partial load with no option to turn them off. Secondly, an improved ENMPC formulation with complementarity constraints (ENMPC+CC 1) is included where $\eta^\text{gtg}(t)$ is directly optimised where the controller can turn off each GTG due to the use of complementarity constraints. Lastly another ENMPC formulation with complementarity constraints (ENMPC+CC 2) where $\eta^\text{gtg}(t)$ is indirectly optimised by constraining the GTG power to be $\bar{P}^\text{gtg}(t) \notin \left(0,\bar{P}^\text{gtg,max}\right)$ is included. Furthermore, another baseline ENMPC which follows \eqref{eq:3-NLP-Formulation-General} where $\eta^\text{gtg}(t)$ is omitted from the cost is included to understand the case where efficiency is not optimised at all (i.e., where the OHPS only aims at meeting some reference with no regards to the optimality of the GTGs). See Table \ref{tb:4-method-names} for the identifiers.
\begin{table}
\begin{center}
\caption{Identifiers of the methods.}\label{tb:4-method-names}
\begin{tabular}{P {1.5cm}||P {6.175 cm}}
\hline
Identifier & Method \\\hline\Tstrut\Bstrut
1 & Baseline ENMPC 1 where $\eta^\text{gtg}_j(t)$ is omitted \\\Tstrut\Bstrut
2 & Baseline ENMPC 2 where $\eta^\text{gtg}_j(t)$ is included\\\Tstrut\Bstrut
3 & CC-ENMPC 1 ($\eta^\text{gtg}_j(t)$ is directly optimised)\\\Tstrut\Bstrut
4 & CC-ENMPC 2 ($\eta^\text{gtg}_j(t)$ is indirectly optimised)\\\hline
\end{tabular}
\end{center}
\end{table}

\subsection{Simulation environment and variables}\label{subsec:4-simenv}

A computer with an Apple M2 is used to simulate the controllers for this study. The controllers are formulated with open-source software with Python and CasADI \citep{Andersson:2019-Casadi}, utilising a multiple shooting approach \citep{Bock:1984-Multiple-Shooting} with Euler's method for integration, and solved with IPOPT \citep{Wachter:2006-IPOPT} using mumps as the linear solver. The simulation study length is set to a day (24 hours) with a time step of 150 s, where the plant and the controller assume no plant-model mismatch, and where the controllers are recomputed every time step with a {prediction horizon} of 120 time steps = 5 hours. The demand to be met is defined based on a Gaussian random variable $\mathcal{N}$ that changes its value every hour based on the magnitude of $P^\text{wtg}(t)$\vspace*{-3 mm} 

\begin{align}
    P^\text{demand}(t) = &P^\text{wtg}(t) + \sum_{{j}=1}^\text{J}0.65P^\text{gtg,max}_j\label{eq:4-power-demand}\\ &+ \mathcal{N}\left(0,0.85P^\text{wtg}(t)\right) \nonumber,
\end{align}

where $P^\text{demand}(t)$ is clipped to make sure that it can be met with the combined power systems at their maximum power output, but also to make sure that the problem is feasible for when the GTGs have to be operated at minimum partial load $P^\text{gtg,min}$ for the methods which do not employ complementarity constraints as these methods cannot turn off the GTGs during operation

\begin{align}
        \sum_{{j}=1}^\text{J}P^\text{gtg,min}_j\leq P^\text{demand}(t) \leq P^\text{ohps,max},\label{eq:4-power-demand-clip}
\end{align}

where $P^\text{ohps,max} = \sum_{{j}=1}^\text{J}P^\text{gtg,max}_j +P^\text{wtg,max} + P^\text{bat,max}$. To highlight the effect of complementarity constraints,  $P^\text{demand}(t)$ and $P^\text{wtg}(t)$ are assumed to be known for the whole simulation study where the hourly average wind speed in this simulation study is obtained from RenewableNinja \citep{Staffel:2016-Renewable-ninja1} around the HyWind Tampen area on 07.06.2019 at the Northern Continental Shelf which is also used to scale the power curve to a maximum value of $P^\text{wtg,max} =$ 88 MW with a static power curve based on real data from a wind farm in Denmark \citep{10366870}. This paper considers an OHPS with 3 LM2500 GTGs in the cluster, each modified so that the maximum power outputs $P^\text{gtg,max}_j$ are 55 MW, 30 MW, and 15 MW, with a minimum partial load requirement at 35\% \citep{GE:2022-factsheet}, and a battery with minimum and maximum power output $P^\text{bat,max}$ and $P^\text{bat,min}$ at $\pm$ 80 MW. Initial system state and input are set to ${x}(0) =$ [1e-3, 1e-3, 1e-3, 1e-3, 1e-3, 1e-3, 70]$^\top$ and ${u}(0) =$ [1e-3, 1e-3, 1e-3, 1e-3]$^\top$ where each method is tuned by trial and error.

\begin{figure*}
    \centering
    {\includegraphics[width=1.75\columnwidth, trim= 0.55cm 0.8cm 0.98cm 0cm, clip]{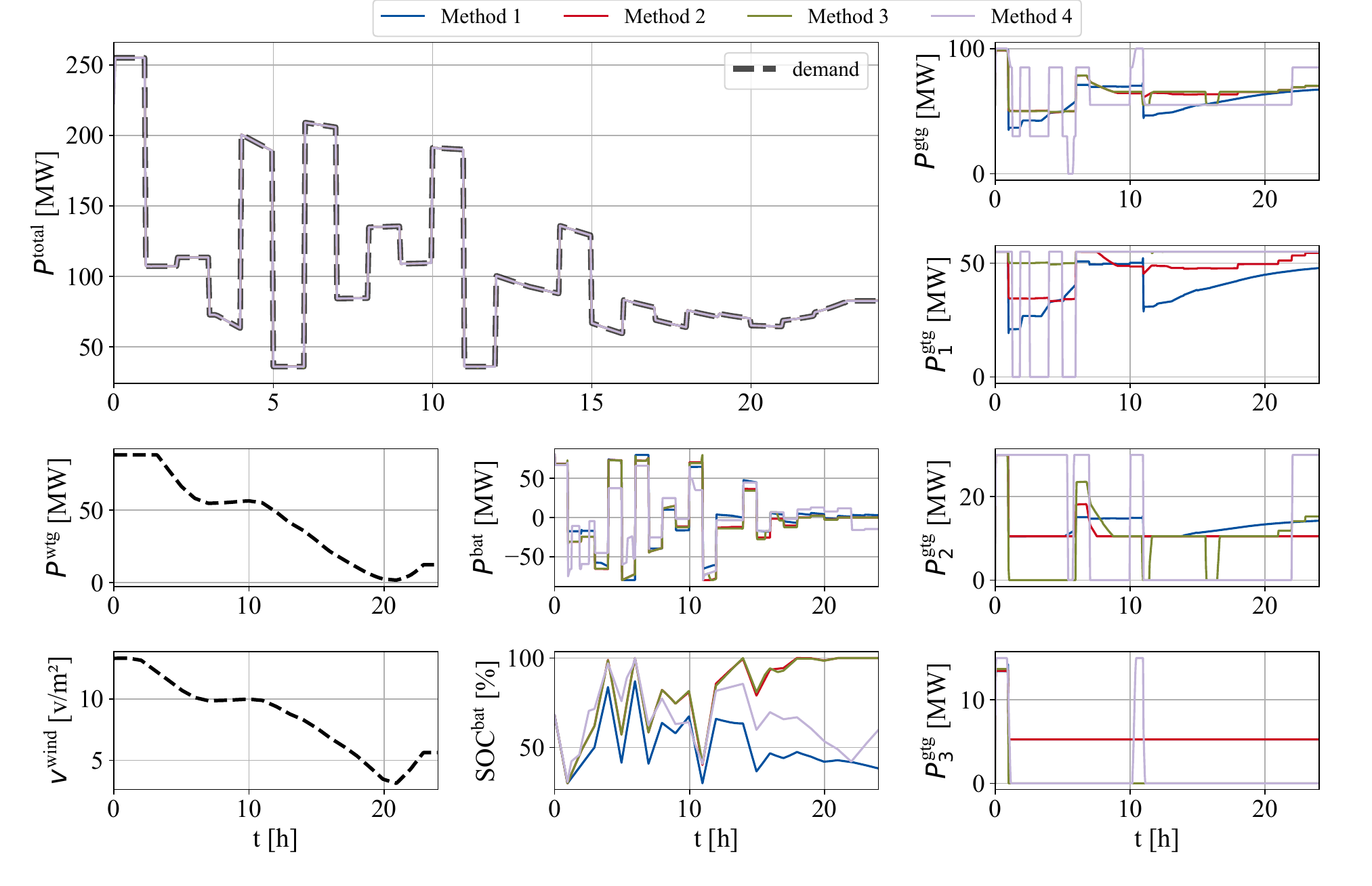}}
    \caption{Time profiles of the total power $P_\text{total}$, GTG power $P_\text{gtg}$, WTG power $P_\text{wtg}$, battery power $P_\text{bat}$, and battery state of charge $\text{SOC}_\text{bat}$ from the different methods, given current wind speed $\text{v}_\text{wind}$. The stipulated black lines are the quantities given for this specific simulation study, such as the wind speed and the demand.}
    \label{fig:4-2-single-results}
\end{figure*}

\subsection{Key performance indicators}\label{subsec:4-key-performance-indicator}
Three key performance indicators which are relevant to control objective 3 are used for the subsequent analysis
\begin{enumerate}
    \item \label{itm:first_kpi} How much power is produced? - $P^\text{gtg}$
    \item \label{itm:second_kpi} How efficient is the produced power (average)? - $\eta^\text{gtg}$
    \item \label{itm:third_kpi} How much emission is released? - $\text{GHG}^\text{gtg}$
\end{enumerate}
which are computed relative to the absolute baseline ENMPC formulation (method 1)\vspace*{-3 mm} 

\begin{equation}
    E_\%^\text{gtg} = \sum_{{j}=1}^\text{J}\int\frac{ P^\text{gtg}_\text{method,i}(t)}{P^\text{gtg}_\text{method,1}(t)}\mathrm{d}t\cdot 100 - 100\label{eq:4-gas-usage-percentage},
\end{equation}
\begin{equation}
    \text{GHG}^\text{gtg}_\% = \sum_{{j}=1}^\text{J}\int\frac{ \text{GHG}^\text{gtg}_\text{method,i}(t)}{\text{GHG}^\text{gtg}_\text{method,1}(t)}\mathrm{d}t\cdot 100 - 100\label{eq:4-gas-efficiency-percentage},
\end{equation}
\begin{equation}
    \bar{\eta}^\text{gtg} = {\int\sum_{{j}=1}^\text{J}\frac{1}{P_j^{\%}}{\eta}^\text{gtg}_j(t)}\mathrm{d}t\cdot 100 - 100\label{eq:4-GHG-usage-percentage},
\end{equation}
where $\sum_{{j}=1}^\text{J} P_j^{\%} = 1$ with $P^{\%}_j =\frac{P^\text{gtg}_j}{\sum_{{j}=1}^\text{J} P^\text{gtg}_j}$ and $i \in [1,4]$. 

The idea is that each subsequent ENMPC formulation after method 1 results in improvement, as they all focus on increasing fuel efficiency. Thus, it is easier to capture this improvement in relative values.

\begin{table}
\begin{center}
\caption{Key performance indicators.}\label{tb:4-single-comparison}
\begin{tabular}{P {1.4 cm}||P {1.85 cm} P {1.85 cm} P {1.85 cm}}
\hline\Tstrut\Bstrut
Method & $E_\%^\text{gtg}$ [\%] & $\bar{\eta}^\text{gtg} [\%]$ & $\text{GHG}^\text{gtg}_\%$ [\%] \\\hline\Tstrut\Bstrut
1 &  0.00 &  43.39 & 0.00 \\\Tstrut\Bstrut
2 &  7.62 & 45.08 & 4.71 \\\Tstrut\Bstrut
3 &  7.56 & 48.49 & -3.25 \\\Tstrut\Bstrut
4 & 2.49  & 50.54 & -11.96 \\\hline
\end{tabular}
\end{center}
\end{table}

\subsection{Simulation study results - Reference tracking}\label{subsec:4-reference tracking}

This subsection examines the proposed methods where the results from applying methods 1-4 for controlling GTGs inside of an OHPS can be seen in Fig. \ref{fig:4-2-single-results}, which shows the time profiles of each power system. Without looking at the GTGs, one can see how $P^\text{wtg}(t)$ follows $\text{v}^\text{wind}(t)$ due to the static wind power curve which is used. Additionally, the power demand also follows $\text{v}^\text{wind}(t)$ due to \eqref{eq:4-power-demand}. Here, the controllable variables to meet the power demand are the GTG powers $P^\text{gtg}_j(t)$ ($P^\text{gtg}$ is here the summed GTG power over the cluster) and the battery power $P^\text{bat}(t)$, where the SOC$^\text{bat}(t)$ is included to help the subsequent analysis. Additionally, the key performance indicators for each method have been collected and shown in Table \ref{tb:4-single-comparison}. The first thing to notice from Fig. \ref{fig:4-2-single-results} is that methods 1 and 2 do not allow the GTGs to be turned off and thus have to operate between 35\% and 100\% partial load. Method 3 and 4, on the other hand, can be turned off when required, with method 4 being constrained to either be at 100\% partial load (maximum power output) or turned off (at 0\%) in contrast to method 3, which can operate continuously between 35\% and 100\%. Something else to notice is that methods 2 and 3 end the simulation run with a fully charged battery. This is achieved by using the GTGs which can be confirmed by looking at Table \ref{tb:4-single-comparison} which shows that methods 2 and 3 use each GTG more than methods 1 and 4. An explanation behind this increased usage can be derived from looking at the cost function of methods 2 and 3, which focuses on maximising the GTG fuel efficiency $\eta^\text{gtg}(t)$. Increasing $\eta^\text{gtg}(t)$ can be achieved by operating the GTGs at a higher partial load. Thus, to maximise $\bar{\eta}^\text{gtg}$, methods 2 and 3 increase their GTG power output by storing excessive GTG power usage with the battery until fully charged. With this, methods 2 and 3 increase the efficiency from method 1, which does not optimise $\eta^\text{gtg}(t)$ at all by 1.69\% and 5.10\%, which increases the GHG for method 2 by 4.71\% in contrast to method 3, where total GHG is decreased by 3.25\%. The increased GHG in method 2 is due to the increased usage of GTGs, while the decreased GHG in method 3 compared to methods 1 and 2 is due to the more efficient operation of the GTGs as a consequence of the use of complementarity constraints, which allows method 3 to completely turn of each GTG instead of letting them run at low efficiency.

\begin{table}
\begin{center}
\caption{Average Computational cost of fully solving the different methods at each time step.}\label{tb:4-nominal-comparison-cpu}
\begin{tabular}{ P{1.74 cm}  P{1.74 cm}  P{1.74 cm}  P{1.74 cm} }
\hline\Tstrut\Bstrut
  \underline{Method 1} & \underline{Method 2} & \underline{Method 3} & \underline{Method 4} \\
  0.04 s & 0.14 s & 0.51 s & 1.12 s\\\hline
\end{tabular}
\end{center}
\end{table}


One can see from Table \ref{tb:4-single-comparison} that method 4 increases the efficiency further by 2.03\% compared to method 3, which decreases the GHG emissions further by 8.71\%. The increase in $\bar{\eta}^\text{gtg}$ is, in contrast, not due to just increasing the GTG power and using the battery proactively to store excess power, but rather by turning each GTG on or off as shown in Fig. \ref{fig:4-2-single-results}. Since method 4 relies on such binary behaviour, instead of storing excess power in the battery, the end SOC$^\text{bat}(t)$ is in a similar range as method 1. This increase in performance is at the cost of increased computational cost, see Table \ref{tb:4-nominal-comparison-cpu}. However, these values still show that the proposed controllers are real-time solvable.

\begin{table*}[ht]
\begin{center}
\caption{Fuel efficiency when varying $K^{\mathrm{d}y}_j$ in \eqref{eq:3-cost-function-proposed-ENMPC-second}.}\label{tb:4-single-comparison-different-K}
\begin{tabular}{P {1.3 cm}|| P {1.5 cm}  P {1.5 cm}  P {1.5 cm} P {1.5 cm} P {1.5 cm} P {1.5 cm} P {1.5 cm}}
\hline\Tstrut\Bstrut
& 1e{-6}$K^{\mathrm{d}y}_j$ & 1e{-5}$K^{\mathrm{d}y}_j$ & 1e{-4}$K^{\mathrm{d}y}_j$ & 1e{-3}$K^{\mathrm{d}y}_j$& 1e{-2}$K^{\mathrm{d}y}_j$ & 1e{-1}$K^{\mathrm{d}y}_j$ & 1e{0}$K^{\mathrm{d}y}_j$\\\hline\Tstrut\Bstrut
$\bar{\eta}^\text{gtg} [\%]$& 50.60   & 50.60     & 50.60     & 50.60  & 50.60   &  50.59   &  50.54 \\\hline
\end{tabular}
\end{center}
\end{table*}
\begin{table*}[ht]
\begin{center}
\begin{tabular}{P {1.3 cm}|| P {1.5 cm}  P {1.5 cm}  P {1.5 cm} P {1.5 cm} P {1.5 cm} P {1.5 cm} P {1.5 cm}}
\hline\Tstrut\Bstrut
  & 1e{1}$K^{\mathrm{d}y}_j$& 1e2$K^{\mathrm{d}y}_j$   & 1e3$K^{\mathrm{d}y}_j$  & 1e{$4$}$K^{\mathrm{d}y}_j$& 1e5$K^{\mathrm{d}y}_j$  & 1e6$K^{\mathrm{d}y}_j$  \\\hline\Tstrut\Bstrut
$\bar{\eta}^\text{gtg} [\%]$  &  50.33   & 49.81    &  49.79 & 48.74 &47.56 &44.87\\\hline
\end{tabular}
\end{center}
\end{table*}

\subsection{Simulation study results - Effect of regularisation}\label{subsec:4-regularisation-effect}

\begin{figure}[ht]
    \centering
    {\includegraphics[width=.95\columnwidth, trim= 0.35cm 0.4cm 0.28cm 0.cm, clip]{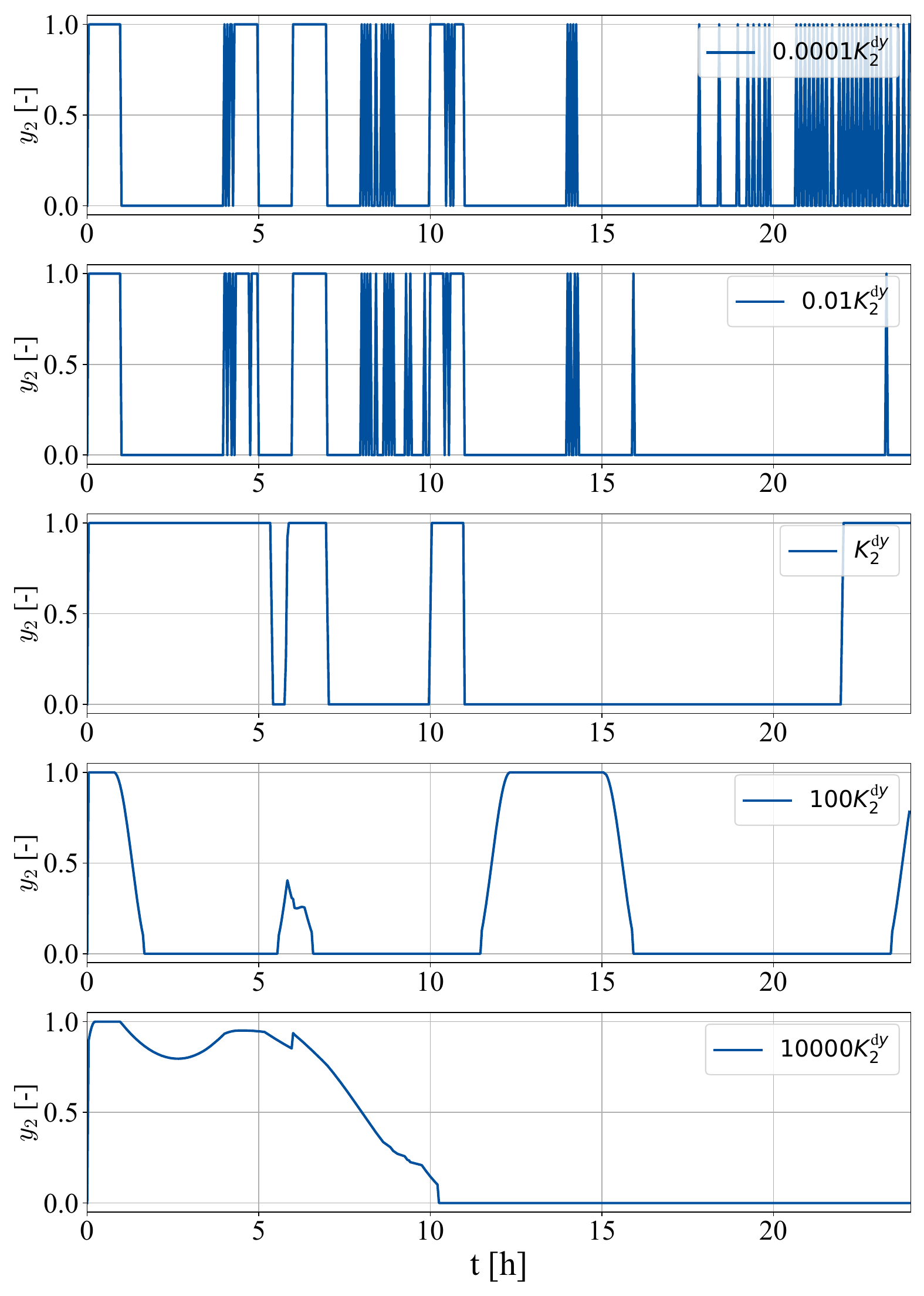}}
    \caption{Time profiles of the second semi-continous variable $y_2$ given different values for $K^{\mathrm{d}y}_j$ in \eqref{eq:3-cost-function-proposed-ENMPC-second}.}
    \label{fig:4-2-regularisation-effect}
\end{figure}

The regularisation term $K^{\mathrm{d}y}_j$ in the complementarity-constrained ENMPC (method 4) plays a vital role in the behaviour of the resulting controller since it directly translates to whether $P_{j}^\text{gtg}$ is at its maximum load, or completely turned off. As can be derived from \eqref{eq:3-cost-function-proposed-ENMPC-second}, $K^{\mathrm{d}y}_j$ is used in $\mathrm{d}y_j(t)^\top  K^{\mathrm{d}y}_j\mathrm{d}y_j(t)$ to minimise the amount of switching between the semi-continuous states of $y_j(t)$ which represents the on/off status of the GTGs. In practice, reducing switching is desirable as it can lead to less maintenance and strain on the equipment. The effect of decreasing and increasing $K^{\mathrm{d}y}_j$ can be seen in Fig. \ref{fig:4-2-regularisation-effect} and Table \ref{tb:4-single-comparison-different-K} under the same simulation environment and wind speed data as the previous study ($y_1$ and $y_3$ are omitted in Fig. \ref{fig:4-2-regularisation-effect} to remove clutter). 

As expected, Fig. \ref{fig:4-2-regularisation-effect} shows that reducing $K^{\mathrm{d}y}_j$ results in more switching between the on and off states of the GTGs which leads to high efficiency as seen in Table \ref{tb:4-single-comparison-different-K}. According to \cite{GE:2022-factsheet}, the LM2500 GTGs can ramp up to 30 - 60 MW/min. Therefore, having a relatively small value for $K^{\mathrm{d}y}_j$ is feasible given that one does not care about the excessive switching, which may lead to stress and fatigue on the turbine. On the other hand, setting $K^{\mathrm{d}y}_j$ too high can result in a more continuous dynamic for $y_j(t)$, as observed in Fig. \ref{fig:4-2-regularisation-effect} which decreases the efficiency as seen in Table \ref{tb:4-single-comparison-different-K}. This behaviour is due to the design choice made in this paper, which implements complementarity constraints as a cost function term in the cost function. This choice allows for a more feasible problem while not entirely constraining $y_j(t)$ to be binary. As $K^{\mathrm{d}y}_j$ becomes too large, the term $\mathrm{d}y_j(t)^\top K^{\mathrm{d}y}_j\mathrm{d}y_j(t)$ directly competes with the complementarity constraints, which can cause the complementarity constraint to break down. One possible solution is to implement complementarity constraints as a constraint instead of as a term in the cost function. However, this approach may lead to numerical difficulties unless relaxed \citep{Hoheisel:2011-relaxation-of-complementarity-constraints}.

\section{Concluding remark}\label{sec:5-conclusion}

This paper proposes two ENMPC formulations for maximising GT fuel efficiency in gas-balanced hybrid power systems with complementarity constraints. One method optimises the GT fuel efficiencies directly, while the other optimises the efficiency indirectly. It can be argued that the latter approach is a special case of the former, where the weight on the efficiency in the cost function is infinite. Simulation results suggest that the latter performs better in terms of fuel efficiency as the direct approach results in a fully charged battery, which offers no future flexibility for storing excess green energy from the wind power system. Another disadvantage of the direct approach arises when one also considers that battery power may leak over time \citep{Ryu:2006-Self-discharge-Lithium-sulfur-batteries}. Future work should investigate modifications of these two proposed formulations. For example, penalising the GHG emissions instead of the efficiency may solve the disadvantages of the direct approach and may be more meaningful when operating GTs, as the ultimate goal is to reduce emissions.

Additionally, future work should expand on this case study by considering wind speed and power demand forecasts at a higher granularity, which, in practice, can be uncertain. A complete framework where power demand is met while operating each GT efficiently despite unreliable forecasts is paramount for actual implementation in real life.  Especially with the recent interest in offshore energy hubs \citep{Zhang:2022-Offshore-Energy-Hubs} for applications such as maritime transport, aquaculture, and green hydrogen production \citep{Mikkola:2018-Multi-platform, Gea:2021-OffshoreHydrogren}, but also since GTGs are expected to be the conventional generator of choice to meet variable and unpredicted changes in net demand for balancing purposes \citep{Baldick:2014-flexibility-with-gas}.

\bibliography{references}

\begin{thebibliography}{23}
\providecommand{\natexlab}[1]{#1}
\providecommand{\url}[1]{\texttt{#1}}
\expandafter\ifx\csname urlstyle\endcsname\relax
  \providecommand{\doi}[1]{doi: #1}\else
  \providecommand{\doi}{doi: \begingroup \urlstyle{rm}\Url}\fi

\bibitem[Andersson et~al.(2019)Andersson, Gillis, Horn, Rawlings, and
  Diehl]{Andersson:2019-Casadi}
J.~Andersson, J.~Gillis, G.~Horn, J.~Rawlings, and M.~Diehl.
\newblock {CasADi} -- {A} software framework for nonlinear optimization and
  optimal control.
\newblock \emph{Mathematical Programming Computation}, 11\penalty0
  (1):\penalty0 1--36, 2019.

\bibitem[Baldick(2014)]{Baldick:2014-flexibility-with-gas}
R.~Baldick.
\newblock Flexibility and availability: Can the natural gas supply support
  these needs?
\newblock \emph{IEEE Power and Energy Magazine}, 12\penalty0 (6):\penalty0
  104--101, 2014.

\bibitem[Baumrucker et~al.(2008)Baumrucker, Renfro, and
  Biegler]{Baumrucker:2008-MPEC-Problem-formulations}
B.~Baumrucker, J.~Renfro, and L.~Biegler.
\newblock Mpec problem formulations and solution strategies with chemical
  engineering applications.
\newblock \emph{Computers \& Chemical Engineering}, 32\penalty0 (12):\penalty0
  2903--2913, 2008.

\bibitem[Biegler(2010)]{Biegler:2010-NLP-book}
L.~T. Biegler.
\newblock \emph{Nonlinear Programming}.
\newblock Society for Industrial and Applied Mathematics, 2010.

\bibitem[Bock and Plitt(1984)]{Bock:1984-Multiple-Shooting}
H.~Bock and K.~Plitt.
\newblock A multiple shooting algorithm for direct solution of optimal control
  problems.
\newblock \emph{9th IFAC World Congress}, 17\penalty0 (2):\penalty0 1603--1608,
  1984.

\bibitem[{GE}(2022)]{GE:2022-factsheet}
{GE}.
\newblock {LM2500 product specifications, GEA32937B}, 2022.

\bibitem[Gea-Bermúdez et~al.(2023)Gea-Bermúdez, Bramstoft, Koivisto, Kitzing,
  and Ramos]{Gea:2021-OffshoreHydrogren}
J.~Gea-Bermúdez, R.~Bramstoft, M.~Koivisto, L.~Kitzing, and A.~Ramos.
\newblock Going offshore or not: Where to generate hydrogen in future
  integrated energy systems?
\newblock \emph{Energy Policy}, 174:\penalty0 113382, 2023.

\bibitem[Hoang et~al.(2022)Hoang, Knudsen, and Imsland]{Hoang:2022-NMPC-OHPS}
K.~Hoang, B.~Knudsen, and L.~Imsland.
\newblock Hierarchical nonlinear model predictive control of offshore hybrid
  power systems.
\newblock \emph{IFAC-PapersOnLine}, 55\penalty0 (7):\penalty0 470--476, 2022.

\bibitem[Hoang et~al.(2023{\natexlab{a}})Hoang, Rugstad~Knudsen, and
  Imsland]{Hoang:2023-M-ENMPC-OHPS}
K.~T. Hoang, B.~Rugstad~Knudsen, and L.~Imsland.
\newblock Reference optimisation of uncertain offshore hybrid power systems
  with multi-stage nonlinear model predictive control.
\newblock In \emph{2023 American Control Conference (ACC)}, pages 1251--1257,
  2023{\natexlab{a}}.

\bibitem[Hoang et~al.(2023{\natexlab{b}})Hoang, Thilker, Knudsen, and
  Imsland]{10366870}
K.~T. Hoang, C.~A. Thilker, B.~R. Knudsen, and L.~Imsland.
\newblock Probabilistic forecasting-based stochastic nonlinear model predictive
  control for power systems with intermittent renewables and energy storage.
\newblock \emph{IEEE Transactions on Power Systems}, pages 1--12,
  2023{\natexlab{b}}.

\bibitem[Hoheisel et~al.(2011)Hoheisel, Kanzow, and
  Schwartz]{Hoheisel:2011-relaxation-of-complementarity-constraints}
T.~Hoheisel, C.~Kanzow, and A.~Schwartz.
\newblock Theoretical and numerical comparison of relaxation methods for
  mathematical programs with complementarity constraints.
\newblock \emph{Mathematical Programming}, 137:\penalty0 1--32, 02 2011.

\bibitem[Jiang et~al.(2021)Jiang, Yuan, Zhang, Bai, Li, Chen, and
  Li]{JIANG2021106460}
T.~Jiang, C.~Yuan, R.~Zhang, L.~Bai, X.~Li, H.~Chen, and G.~Li.
\newblock Exploiting flexibility of combined-cycle gas turbines in power system
  unit commitment with natural gas transmission constraints and reserve
  scheduling.
\newblock \emph{International Journal of Electrical Power \& Energy Systems},
  125:\penalty0 106460, 2021.

\bibitem[K{\"o}ppe(2012)]{Koppe:2012-Complexit-of-Nonlinear-Mixed-integer}
M.~K{\"o}ppe.
\newblock On the complexity of nonlinear mixed-integer optimization.
\newblock In \emph{Mixed Integer Nonlinear Programming}, pages 533--557, New
  York, NY, 2012. Springer.

\bibitem[Mikkola et~al.(2018)Mikkola, Heinonen, Kankainen, Hekkala, and
  Kurkela]{Mikkola:2018-Multi-platform}
E.~Mikkola, J.~Heinonen, M.~Kankainen, T.~Hekkala, and J.~Kurkela.
\newblock Multi-platform concepts for combining offshore wind energy and fish
  farming in freezing sea areas: Case study in the gulf of bothnia.
\newblock In \emph{ASME 2018 37th International Conference on Ocean, Offshore
  and Arctic Engineering}, volume~6, Oct. 2018.

\bibitem[Nirbito et~al.(2020)Nirbito, Budiyanto, and
  Muliadi]{Nirbito:2020-GTG-performance-Boat}
W.~Nirbito, M.~A. Budiyanto, and R.~Muliadi.
\newblock Performance analysis of combined cycle with air breathing derivative
  gas turbine, heat recovery steam generator, and steam turbine as lng tanker
  main engine propulsion system.
\newblock \emph{Journal of Marine Science and Engineering}, 8:\penalty0 726,
  2020.

\bibitem[{Norwegian Petroleum Directorate}(2019)]{NPD:2019-ResourceReport}
{Norwegian Petroleum Directorate}.
\newblock {Resource Report}, 2019.

\bibitem[Pan et~al.(2016)Pan, Guan, Watson, and Wang]{7112194}
K.~Pan, Y.~Guan, J.-P. Watson, and J.~Wang.
\newblock Strengthened milp formulation for certain gas turbine unit commitment
  problems.
\newblock \emph{IEEE Transactions on Power Systems}, 31\penalty0 (2):\penalty0
  1440--1448, 2016.

\bibitem[{R. Pavri, G.D. Moore}(2001)]{GE:2001-GT-emissions-and-control}
{R. Pavri, G.D. Moore}.
\newblock { Gas turbine emissions and control. GE Reference Library, GER-
  4211}, 2001.

\bibitem[Rawlings et~al.(2017)Rawlings, Mayne, and
  Diehl]{Rawlings:2017-MPCBook}
J.~Rawlings, D.~Mayne, and M.~Diehl.
\newblock \emph{Model Predictive Control: Theory, Computation, and Design}.
\newblock Nob Hill Publishing, Jan. 2017.

\bibitem[Ryu et~al.(2006)Ryu, Ahn, Kim, Ahn, Cho, and
  Nam]{Ryu:2006-Self-discharge-Lithium-sulfur-batteries}
H.~Ryu, H.~Ahn, K.~Kim, J.~Ahn, K.~Cho, and T.~Nam.
\newblock Self-discharge characteristics of lithium/sulfur batteries using
  tegdme liquid electrolyte.
\newblock \emph{Electrochimica Acta}, 52\penalty0 (4):\penalty0 1563--1566,
  2006.

\bibitem[Staffell and Pfenninger(2016)]{Staffel:2016-Renewable-ninja1}
I.~Staffell and S.~Pfenninger.
\newblock Using bias-corrected reanalysis to simulate current and future wind
  power output.
\newblock \emph{Energy}, 114:\penalty0 1224--1239, 2016.

\bibitem[Wächter and Biegler(2006)]{Wachter:2006-IPOPT}
A.~Wächter and L.~Biegler.
\newblock On the implementation of an interior-point filter line-search
  algorithm for large-scale nonlinear programming.
\newblock \emph{Mathematical programming}, 106:\penalty0 25--57, 03 2006.

\bibitem[Zhang et~al.(2022)Zhang, Tomasgard, Knudsen, Svendsen, Bakker, and
  Grossmann]{Zhang:2022-Offshore-Energy-Hubs}
H.~Zhang, A.~Tomasgard, B.~R. Knudsen, H.~G. Svendsen, S.~J. Bakker, and I.~E.
  Grossmann.
\newblock Modelling and analysis of offshore energy hubs.
\newblock \emph{Energy}, 261:\penalty0 125219, 2022.

\end{thebibliography}




\end{document}